# Superparamagnetic nanoparticle ensembles

O. Petracic

*Institute of Experimental Physics/Condensed Matter Physics, Ruhr-University Bochum, 44780*

*Bochum, Germany*

**Abstract:** Magnetic single-domain nanoparticles constitute an important model system in magnetism. In particular ensembles of superparamagnetic nanoparticles can exhibit a rich variety of different behaviors depending on the inter-particle interactions. Starting from isolated single-domain ferro- or ferrimagnetic nanoparticles the magnetization behavior of both non-interacting and interacting particle-ensembles is reviewed. A particular focus is drawn onto the relaxation time of the system. In case of interacting nanoparticles the usual Néel-Brown relaxation law becomes modified. With increasing interactions modified superparamagnetism, spin glass behavior and superferromagnetism are encountered.

## 1. Introduction

Nanomagnetism is a vivid and highly interesting topic of modern solid state magnetism and nanotechnology [1-4]. This is not only due to the ever increasing demand for miniaturization, but also due to novel phenomena and effects which appear only on the nanoscale. That is e.g. superparamagnetism, new types of magnetic domain walls and spin structures, coupling phenomena and interactions between electrical current and magnetism (magneto resistance and current-induced switching) [1-4]. In technology nanomagnetism has become a crucial commercial factor. Modern magnetic data storage builds on principles of nanomagnetism and this tendency will increase in future. Also other areas of nanomagnetism are commercially becoming more and more important, e.g. for sensors [5] or biomedical applications [6]. Many potential future applications are investigated, e.g. magneto-logic devices [7], [8], photonic systems [9, 10] or magnetic refrigeration [11, 12].

In particular magnetic nanoparticles experience a still increasing attention, because they can serve as building blocks for e.g. data storage media, spintronic devices, photonic or biomedical systems. The term 'magnetic nanoparticles' (sometimes called 'nanoclusters', 'nanocrystals', 'fine particles' or 'nanobeads') refers to more or less spherical particles with diameters in the range $d \approx 3$ to 30 nm from a ferro-, ferri- or antiferromagnetic material [13-17]. The investigations on ferro- or ferrimagnetic particles constitutes the dominant branch in research and applications, whereas antiferromagnetic nanoparticles form a more specific but still equally interesting topic in nanomagnetism [18-22].

Typical ferromagnetic materials for nanoparticles are all 3d-ferromagnets, i.e. Fe, Co, Ni and its alloys, e.g. CoFe, $Ni_{80}Fe_{20}$ and FePt, various Oxides, e.g. Magnetite $Fe_3O_4$ or Nitrides, e.g. $\varepsilon$-$Fe_3N$. There exist also various ferrimagnets, e.g. Maghemite $\gamma$-$Fe_2O_3$.

## 2. Stoner-Wohlfarth model

The starting point of the discussion is the Stoner-Wohlfarth model, which can be introduced from considering a simplified Hamiltonian for a magnetic system. I.e. the static and quasi-static magnetization, $\mathbf{M}(\mathbf{r})$, of any classical magnetic system follows from the minimization of the free energy and in case of temperatures well below the ordering temperature it basically follows from the minimization of the energy. A simple model ansatz is

$$E = E_J + E_K + E_H + E_d =$$
$$= -J \sum_{<i,j>} \mathbf{S}_i \cdot \mathbf{S}_j - K \sum_i (\hat{\mathbf{k}}_i \cdot \mathbf{S}_i)^2 + g\mu_B \mu_0 \sum_i \mathbf{H} \cdot \mathbf{S}_i + \frac{(g\mu_B)^2 \mu_0}{4\pi} \sum_{<i,j>} \left( \frac{\mathbf{S}_i \cdot \mathbf{S}_j}{\mathbf{r}_{ij}^3} - 3\frac{(\mathbf{S}_i \cdot \mathbf{r}_{ij})(\mathbf{S}_j \cdot \mathbf{r}_{ij})}{\mathbf{r}_{ij}^5} \right),$$
(1)

where $E_J$ is the exchange energy, $E_K$ the anisotropy energy. $E_H$ is the Zeeman energy in an applied magnetic field $\mathbf{H}$ and $E_d$ the dipolar coupling energy between all spins (or in general moments) at a distance $\mathbf{r}_{ij}$, $g$ the relevant g-factor and $\mu_B$ the Bohr magneton. Often several exchange and anisotropy contributions can exist in one system. Here for simplicity only one exchange and one uniaxial anisotropy term with anisotropy constant $K$ and anisotropy axis $\hat{\mathbf{k}}$ is given. The 'spin vector' $\mathbf{S}_i$ has to be understood as a general *dimensionless* 'space holder'. This can be the spin quantum number $\mathbf{S}_i$, e.g. $\mathbf{S}_i = (0, 0, \pm 1/2)$ in the case of the Ising-model, the total angular momentum $\mathbf{J}_i$ or in an often employed semi-classical approximation the magnetostatic moment $\mathbf{m}_i/\mu_B$ as a classical vector.

It is well known that the dipolar term is responsible for the occurrence of domains, since it creates together with the shape of the specific system demagnetization fields [23-26]. The system minimizes its stray field energy by fragmentation into domains. Fig. 1 shows two examples of spin structures of a spherical ferromagnetic nanoparticle, where in (a) only one domain ('single-domain state') occurs with large stray fields, and in (b) a particle with two domains and thus reduced stray fields. The domain wall is marked by the dotted line.

On the other hand the introduction of domain walls costs a surface energy density of the order of $\sqrt{AK}$, where $A \propto J$ is the 'exchange stiffness' [23]. One can easily estimate the critical radius of, e.g. a spherical particle, below which the gain in reduction of stray fields is less then the cost of introducing a domain wall [23, 27, 28], i.e.

$$R_c = \frac{36\sqrt{AK}}{\mu_0 M_s^2}. \qquad (2)$$

Therefore, a particle with radius $r < R_c$ will prefer to stay in a 'single-domain state' [Fig. 1(a)]. Values for $R_c$ are in the order of ~10nm... 100nm [28], e.g. $R_c$(Co) =34nm and $R_c$(Fe$_3$O$_4$) = 49nm.

When trying to reverse the magnetization of a single-domain particle basically three possible scenarios exist: reversal by (i) 'curling', (ii) 'buckling' or (iii) by 'coherent rotation' [27, 28]. The third describes the simultaneous rotation of all moments in the particle in an 'unison' fashion. This case is a convenient model case, because it describes a system with one large moment associated to one nanoparticle, $\mathbf{m}_{NP}$ with $|\mathbf{m}_{NP}| \approx M_s V \approx$ const being analogous to an atomic moment. Due to the huge values of $|\mathbf{m}_{NP}| \sim 1000\mu_B$ it is usually referred to a as 'superspin'. In this case the expression for the energy in Eq. (1) reduces to

$$E_{NP} = -KV(\hat{\mathbf{k}} \cdot \hat{\mathbf{m}}_{NP})^2 - \mu_0 M_s V \mathbf{H} \cdot \hat{\mathbf{m}}_{NP}, \tag{3}$$

where $K$ is an effective uniaxial anisotropy constant including both the magnetocrystalline and demagnetizing contribution and $V$ the volume of the nanoparticle. Considering only rotation-symmetric cases this expression can be simplified to

$$E_{SW} = KV \sin^2\phi - \mu_0 H M_s V \cos(\theta - \phi), \tag{4}$$

with $\phi$ being the angle between magnetic moment and anisotropy axis and $\theta$ the angle between applied field and the anisotropy axis. This expression (4) is the basis of the *Stoner-Wohlfarth model* [23-28]. A typical rotation-symmetric system is realized in prolate spheroid nanoparticles as depicted in Fig. 2 (a). When crystal and surface anisotropies are negligible then the anisotropy energy is governed by the shape anisotropy of the particle. In this simple model at zero field there will be two degenerate energy minima at $\phi = 0°$ and $180°$ [Fig. 2 (b)]. In case of an applied field, the energy minima shift vertically so that one local and one absolute minimum arises.

## 3. Superparamagnetism

At $H = 0$ the two minima are separated by an energy barrier of height $\Delta E = KV$. If $KV \gg k_B T$ then the moment $\mathbf{m}_{NP}$ cannot switch spontaneously. Then, the system behaves like a 'permanent' ferromagnet. However, if the energy barrier is of the order of the thermal energy, $KV \sim k_B T$ or less, then spontaneous switching of the 'superspins' can occur on the timescale of the experiment. In this case one speaks of *superparamagnetic (SPM) nanoparticles*.

The thermally exited fluctuations of the superspin-directions take place with a frequency $f$ or a characteristic *relaxation time* $\tau = (2\pi f)^{-1}$. A quantitative expression for $\tau$ is given by the *Néel-Brown model* reflecting an Arrhenius type of activation law [13-15, 29, 30],

$$\tau = \tau_0 \exp\left(\frac{KV}{k_B T}\right), \tag{5}$$

where the pre-factor $\tau_0 \sim 10^{-9}$ s. In the presence of an applied field this expression is modified to

$$\tau = \tau_0 \exp\left(\frac{\Delta E(H,\theta)}{k_B T}\right), \tag{6}$$

with $\Delta E(H, \theta)$ being the field dependent energy barrier, which can be expressed as [14, 31-33]

$$\Delta E(H,\theta) = \Delta E_0 \left(1 - \frac{H}{H_{SW}^0}\right)^\kappa, \tag{7}$$

with $\kappa = 0.86 + \frac{1.14 H_{SW}^0}{H_a}$, and $H_{SW}^0 = \frac{H_a}{(\sin^{2/3}\theta + \cos^{2/3}\theta)^{3/2}}$,

and $H_a := \frac{2K}{M_s}$.

In Fig. 3 the $1/T$ temperature dependence of the relaxation time according to Eq. (5) is shown in a logarithmic plot for the case $KV/k_B = 315$ K at $H = 0$. The astounding observation is the huge span of time scales covered in the temperature range shown. While $\tau \sim 10^{-9}$ s at $T = 300$ K it increases by 27 orders of magnitude to $\sim 10^{+18}$ s at 5 K.

Therefore, the dynamics of the system is strongly governed by the temperature. At high temperatures the magnetic moments will rapidly fluctuate, whereas at low temperatures they will appear 'blocked'. It is clear that there is a characteristic crossover-temperature, which separates the 'free' from the 'blocked' regime for a given probing time scale of a measurement. This characteristic temperature is called *blocking temperature*, $T_B$, which is given by the temperature at which the time scale of the nanoparticle fluctuations and that of the measurement match, i.e.,

$$\tau_{exp} = \tau(T_B) = \tau_0 \exp\left(\frac{KV}{k_B T_B}\right) \Rightarrow T_B = \frac{KV}{k_B \ln(\tau_{exp}/\tau_0)} \tag{8}$$

In the example presented in Fig. 3 the time scale of a SQUID magnetometry experiment is ~10 s and thus $T_B = 14$ K. Therefore, for $T < T_B$ the superspins will appear blocked or frozen, while for $T > T_B$ they will appear free comparable to a paramagnetic system. One should note that the blocking temperature is not an intrinsic temperature of the system but depends strongly on the measurement time.

This has important consequences for the magnetization behavior of a SPM system. In a hysteresis loop two possible shapes can be observed. I.e. when the time scale to drive through the hysteresis loop, $\tau_{M\text{-}H}$ is larger than the relaxation time of the nanoparticles at a given temperature, i.e., $\tau_{M\text{-}H} \gg \tau(T)$, then the loop appears closed and S-shaped as for a paramagnetic system. However, if $\tau_{M\text{-}H} < \tau(T)$, then a *open* hysteresis loop can be observed with a finite dynamic coercive field, $H_c^*(T, \tau_{M\text{-}H})$ [34], resembling a ferromagnetic system. Note that this can be valid even *above* the blocking temperature of a system, since the magnetization behavior is given by the relative time scales. Therefore, extra care has to be taken when interpreting hysteresis curves of granular systems.

Another type of measurement is to record *M* vs. *T*. Fig. 4 shows a typical example of the zero field cooled (ZFC) and field cooled (FC) magnetization curve as function of *T* from Monte-Carlo simulations of an ensemble of non-interacting superparamagnetic particles with an energy barrier of $KV/k_B = 315$ K and $M_sV/k_B = 1475$ K/T at $\mu_0 H = 0.04$ T and a random distribution of anisotropy axes. The ZFC-curve is measured upon heating in a specified field *after* cooling the sample from a high temperature above the blocking temperature (or generally a transition temperature) down to a low temperature in zero applied field. The FC-curve is measured in the same field subsequently after recording the ZFC curve. It can be measured either upon cooling or upon warming after cooling-in the system in this field.

One finds the typical splitting of the ZFC and FC curves for SPM systems below a splitting temperature $T_s$. Moreover, the ZFC curve exhibits a maximum, which usually defines the (experimental) blocking temperature, $T_B$. It is interesting to note that the value here for $T_B \approx$ 25 K does not match the one from Fig. 3, i.e. 14 K, although the energy barrier has exactly the same value. The reason is that the time scales are different. Therefore it is important to note that the blocking temperature is not an absolute and characteristic temperature, but depends on the probing time scale.

The splitting and the blocking temperature do not need to match in general. Only for the case of identical energy barriers for all particles one finds $T_s = T_B$. For a finite dispersion of sizes (or more general: energy barriers) one finds $T_s > T_B$.

In the context of the next section "Effect of interactions between nanoparticles" it makes sense to define the term *'superparamagnetic system (or ensemble)'* as an ensemble of *non-interacting* SPM nanoparticles. This is necessary in order to distinguish the magnetic behavior of such a system from the behavior of interacting SPM nanoparticles, which can show collective properties or even a phase transition.

## 4. Effect of interactions between nanoparticles

The most relevant interaction in experimental nanoparticle systems is the dipole interaction. Then Eq. 3 becomes:

$$E_i = -K_i V_i (\hat{\mathbf{k}} \cdot \hat{\mathbf{m}}_i)^2 - \mu_0 M_s V_i \mathbf{H} \cdot \hat{\mathbf{m}}_i + \frac{\mu_0 M_s^2}{4\pi} \sum_{\{j\}} V_i V_j \left( \frac{\hat{\mathbf{m}}_i \cdot \hat{\mathbf{m}}_j}{\mathbf{r}_{ij}^3} - 3 \frac{(\hat{\mathbf{m}}_i \cdot \mathbf{r}_{ij})(\hat{\mathbf{m}}_j \cdot \mathbf{r}_{ij})}{\mathbf{r}_{ij}^5} \right) \quad (9)$$

where $E_i$ is the energy and $\mathbf{m}_i$ the magnetic moment of the particle with index $i$. The dipole interaction is dominant in those systems, where the surrounding matrix is both an insulator and diamagnetic. This is the case e.g. in ferrofluids, where the nanoparticles are dispersed in a solvent [35-37] or where they are embedded in a crystalline or amorphous solid e.g. $Al_2O_3$ [38-40] or $SiO_2$ [41]. In case of a conducting matrix also RKKY interactions can play a role [23-26] and, eventually, for a strongly paramagnetic matrix like Al, Cr or Pd one could even think of a mediated interaction via the polarization of the paramagnet by the superspins.

Depending on the strength of the interaction one finds different magnetic behaviors of the ensemble. A useful quantity to characterize and classify the behavior is the relaxation time, $\tau$. One can distinguish the following behaviors [13]:

(i) Superparamagnetism (SPM): $$\tau = \tau_0 \exp\left(\frac{KV}{k_B T}\right), \tag{10}$$

(ii) Modified superparamagnetism $$\tau = \tau_0 \exp\left(\frac{\Delta E^*}{k_B T}\right), \tag{11}$$

(iii) Glass-like freezing: $$\tau = \tau_0 \exp\left(\frac{\Delta E^*}{k_B (T-T_0)}\right) \tag{12}$$

(iv) Superspin glass (SSG): $$\tau = \tau_0^* \left(\frac{T-T_g}{T_g}\right)^{-z\nu} \tag{13}$$

(v) Superferromagnetism (SFM): $$\tau = \tau_0^* \left(\frac{T-T_c}{T_c}\right)^{-z\nu}. \tag{14}$$

In case (i) the ensemble is characterized by the independent, individual behavior of the single particles. This applies also to case (ii), however, here the energy barrier, $\Delta E^*$, is modified by an effective contribution of the inter-particle interactions [13].

With increasing interaction strength one can encounter *collective* behavior. One has to distinguish between the following cases. In case (iii) one finds *glass-like* freezing of the superspins. It is not a true phase transition with critical behavior. The Néel-Brown law is modified by adding a 'glass temperature', $T_0$ in the denominator. This expression is refereed to as the Vogel-Fulcher law (Eq. 12) [13, 43]. For even stronger interactions (viz. closer distances) and if magnetic frustration and spatial randomness is present one can encounter a phase transition into a so-called superspin glass phase below a critical temperature, $T_g$ [case (iv)]. Then, the relaxation time is given by a power law (13) with the critical exponent $z\nu$, where $\nu$ is the critical exponent of the correlation length, $\xi \propto ([T-T_g]/T_g)^{-\nu}$. And the exponent $z$ relates the relaxation time with the correlation length via $\tau \propto \xi^z$ [42]. This case is termed *superspin glass*, since the behavior is completely analogous to canonical spin glasses like $Au_xFe_{1-x}$ [43]. Instead of atomic spins the particle superspins freeze into a spin glass phase below a critical temperature, $T_g$ [13-15, 40, 44-48].

Finally, if the dipolar interactions are strong enough and the spatial arrangement of nanoparticles ordered enough then case (v) can be encountered, where ferromagnetic-like correlations of the superspin moments occur [40, 49-53]. This system is then termed superferromagnet (SFM) [49].

The different behaviors can be roughly discriminated by successful vs. unsuccessful fit to the data. A successful fit means that reasonable values for the fit parameters are found. E.g. a SPM system will yield zero or an unreasonable value for $T_g$ when fitted to a SSG law. In contrast, the data of a SSG system yields unreasonable values for $\tau_0$ when fitted with a SPM law. However, a correct distinction can only be obtained by employing several measurements as described in Ref. [40].

## 5. Interacting ferromagnetic CoFe nanoparticles

A specifically attractive system to study the above mentioned cases are so called 'discontinuous metal insulator multilayers' (DMIMs) [44, 53, 54]. We have studied DMIMs of type $[Co_{80}Fe_{20}(t_{CoFe})/Al_2O_3(3nm)]_N$, which were prepared by sequential Xe ion beam sputtering from $Co_{80}Fe_{20}$ and $Al_2O_3$ targets on a glass substrate. Here $t_{CoFe}$ is the nominal thickness of the CoFe and $N$ the number of repetitions of $Co_{80}Fe_{20}/Al_2O_3$ bilayers. The CoFe does not form a continuous layer, but nanoparticles due to non-wetting on the $Al_2O_3$ layers. Because of the insulating matrix the inter-particle interactions are of magnetic dipolar nature. Fig. 5 shows in (a) a schematic crossection and in (b) a on top transmission electron microscopy (TEM) image of one CoFe/$Al_2O_3$ bilayer [48, 53].

The size of the nanoparticles can be tuned by the nominal thickness, $t_{CoFe}$, of the CoFe layer. We have investigated samples with $0.5 \leq t_{CoFe} \leq 1.8$ nm and numbers of bilayers $1 \leq N$

≤ 10. In case of $t_{CoFe}$ = 0.9 nm one achieves nanoparticles with a diameter of 3 nm and mean distance of 6 nm. In the range $t_{CoFe}$ ≤ 1.2 nm the diameter scales approximately as $(t_{CoFe})^{1/3}$, while the number density stays constant. With increasing $t_{CoFe}$ the inter-particle distance does not change. However, due to the increasing particle diameter and thus increasing magnetic moment also the dipolar interaction strength increases. For $t_{CoFe}$ > 1.2 nm a critical particle height is reached and particle continue to grow only laterally. At $t_{CoFe}$ ≈ 1.8 nm laterally geometric percolation is found, which is accompanied by a crossover from tunnel to metallic electrical conductivity [38].

Therefore it is possible to tune through different regimes of dipolar interaction strengths by choosing $t_{CoFe}$. In fact one can find the following sequence of magnetic behavior depending on $t_{CoFe}$:

$t_{CoFe}$ ≤ 0.5 nm,         SPM,                Ref. [55],

0.5 < $t_{CoFe}$ < 0.8 nm,   modified SPM,       Ref. [55],

0.8 ≤ $t_{CoFe}$ < 1.2 nm,   SSG,                Ref. [40, 44, 53, 55],

1.2 ≤ $t_{CoFe}$ < 1.5 nm,   SFM,                Ref. [40, 52, 53, 56, 57].

This can be also depicted as a magnetic phase diagram. In Fig. 6 the characteristic transition temperatures are shown as function of nominal thickness of a ferromagnetic material, $t_{FM}$ [40]. Increasing nominal thickness implies both increasing particle size and increasing interaction strength.

There are three characteristic transitions lines. I.e. at high temperatures, on finds the bulk Curie temperature of the ferromagnetic material, $T_{c,bulk}$ (broken thin line). For nanoparticles this transition temperature will be reduced due to the finite size effect. Below this temperature there will be ferromagnetic ordering *inside* each particle. However, no inter-particle ordering can be found. Therefore, in analogy to the paramagnetic phase, one can denote this phase as 'SPM phase'. For a given measurement speed one will find a blocking temperature $T_B$ of the *individual* nanoparticles. Experimentally, this value is only accessible for negligible inter-particle interactions. It is possible, however not trivial, to distinguish the finite-size Curie temperature and the blocking temperature of the nanoparticles [59].

Since the nominal thickness, $t_{FM}$, changes also the particle diameter and hence the particle energy barrier, $T_B(t_{FM})$ will be a curve with positive slope (broken thick line). For small enough $t_{FM}$ one finds $T_B \propto (t_{FM})^{1/3}$, because of $T_B \propto KV_{NP}$ and $V_{NP} \propto (t_{FM})^{1/3}$.

The thick solid line marks the transition temperatures of a *collective* state, i.e., either the spin glass temperature of the SSG phase, $T_g$, or the critical temperature of the SFM phase, $T_c$. Hence, on encounters four cases (Fig. 6):

(i) For very small $t_{FM}$ and hence small interaction strength, the ensemble is characterized by individual blocking or modified individual blocking of SPM particles. The ensemble is characterized by the blocking temperature, $T_B$.

(ii) In this region the inter-particle interactions would be sufficient to induce collective order (either SSG or SFM). However, the corresponding transition temperature ($T_g$ or $T_c$, respectively) is *smaller* than the blocking temperature of the individual nanoparticles. Consequently, the collective order is hidden beneath SPM blocking behavior [55].

(iii) For strong enough interactions the transition temperature of a collective state becomes larger than $T_B$ and hence a phase transition can be measured. In case of moderate interaction strength and particle position disorder a SSG state can be found.

(iv) For even stronger interactions and sufficient spatial particle arrangement a SFM is encountered.

The experimental phase diagram of the system $[Co_{80}Fe_{20}(t_{CoFe})/Al_2O_3(3nm)]_{10}$ is depicted in Fig. 7 for the range of nominal thickness 0.9 nm $\leq t_{CoFe} \leq$ 1.4 nm. One finds SSG behavior for $0.8 \leq t_{CoFe} <$ 1.2 nm [40, 44, 53, 55]. For approximately $t_{CoFe} <$ 0.8 nm one encounters modified SPM behavior. SFM behavior is found for $1.2 \leq t_{CoFe} <$ 1.5 nm [40, 52, 53, 56, 57].

Above approximately 1.5 nm geometric percolation starts to set in. Then, the magnetic behavior resembles that of a random ferromagnet [58]. A relatively good correspondence of the experimental to the schematical phase diagram can be found. The three phases SPM, SSG and SFM can be clearly identified.

In summary, the basics of superparamagnetic nanoparticles has been reviewed. In the case when inter-particle interactions become relevant, one finds that the relaxation time $\tau$ is an important quantity to classify the overall behavior of the particle ensemble. Depending on the strength of interaction different cases occur: superparamagnetism, modified superparamagnetism, superspin glass behavior and superferromagnetism. A schematic expected phase diagram with the nominal thickness of the ferromagnetic material is compared to the experimental one of the discontinuous multilayer system $[Co_{80}Fe_{20}(t_{CoFe})/Al_2O_3(3nm)]_{10}$. The phases of superparamagnetism, superspin glass behavior and superferromagnetism can be clearly identified.

**Figure Captions:**

Fig. 1. Two possible spin structures of a spherical ferromagnetic nanoparticle (a) with only one domain (single-domain state) and large stray fields, and (b) a system with two domains separated by a 180° domain wall (dotted line) and thus reduced stray fields.

Fig. 2. (a) Schematics of a prolate nanoparticle in the Stoner-Wohlfarth model. Panel (b) shows a plot of $E_{SW} / 2KV$ as function of the angle $\phi$ for $h = \mu_0 H M_s / 2KV = 0$ (blue curve) und $h = 0.2$ (red curve) with $\theta = 0°$ for simplicity.

Fig. 3. Plot of the relaxation time $\tau = \tau_0 \exp(KV/k_B T)$ vs. $1/T$ for $KV/k_B = 315$ K (blue straight line). The typical time scale of a SQUID-experiment, ~10 s, is marked by the broken line. It intersects the curve $\tau$ vs. $1/T$ at $T = 14$ K thus defining the blocking temperature of the system.

Fig. 4. Magnetization curve after ZFC, $M_{ZFC}(T)$, and after FC, $M_{FC}(T)$, from Monte-Carlo simulations of an ensemble of non-interacting superparamagnetic particles with $KV/k_B = 315$ K and $M_s V/k_B = 1475$ K/T at $\mu_0 H = 0.04$ T. The blocking temperature, $T_B$, is marked by an arrow.

Fig. 5. Schematic crossection (left) and a on-top view TEM image (right) of one CoFe(0.9nm)/Al$_2$O$_3$ bilayer taken from Ref. [48] and [53].

Fig. 6. General schematic phase diagram of a discontinuous ferromagnet-insulator system with the nominal thickness of the ferromagnetic layer as control parameter taken from Ref. [40]. $T_B$ is the blocking temperature of the individual particles, $T_g$ the glass transition temperature in case of a SSG state, $T_c$ the critical temperature of a SFM state and $T_{c,bulk}$ the bulk Curie temperature of the ferromagnetic material. PM denotes the paramagnetic phase. For numbers (i)-(iv) see text.

Fig. 7. Experimental phase diagram of the system $[Co_{80}Fe_{20}(t_{CoFe})/Al_2O_3(3nm)]_{10}$ for the range of nominal thickness $0.9 \leq t_{CoFe} \leq 1.4$ nm taken from Ref. [40].

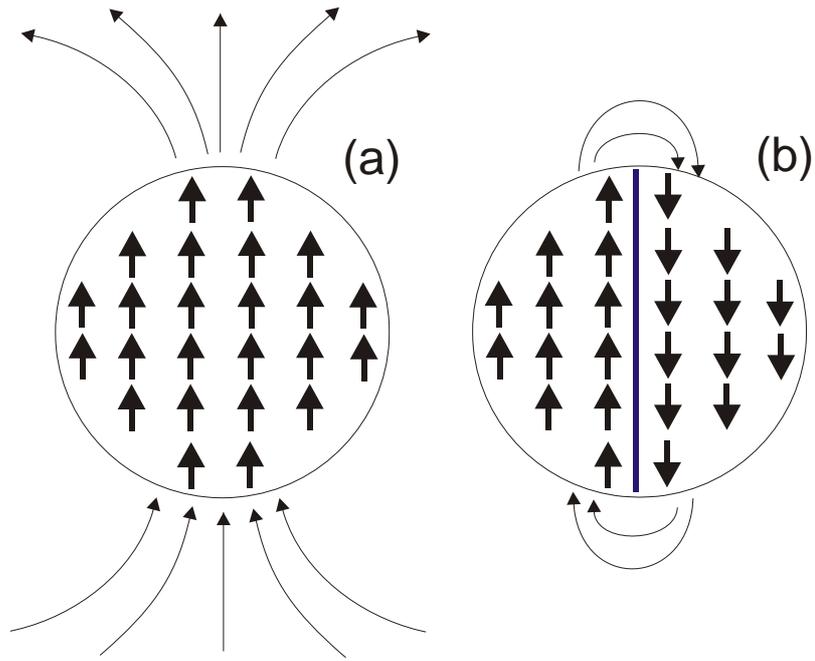

Fig. 1, Petracic et al.

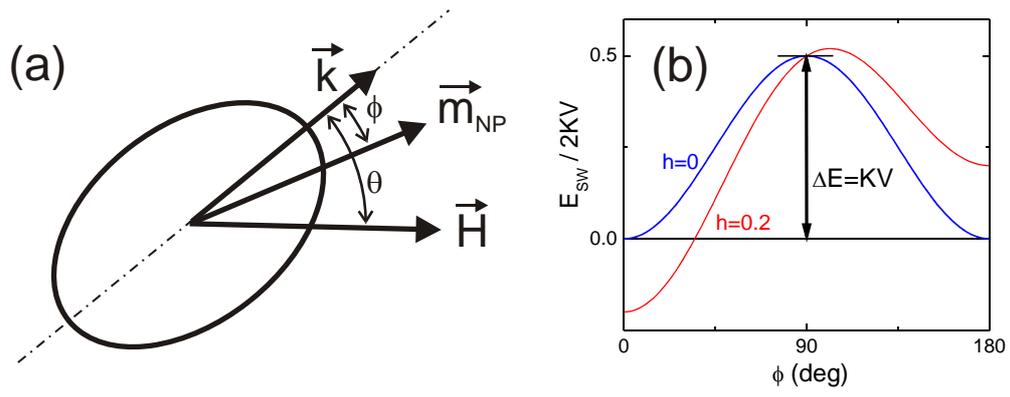

Fig. 2, Petracic et al.

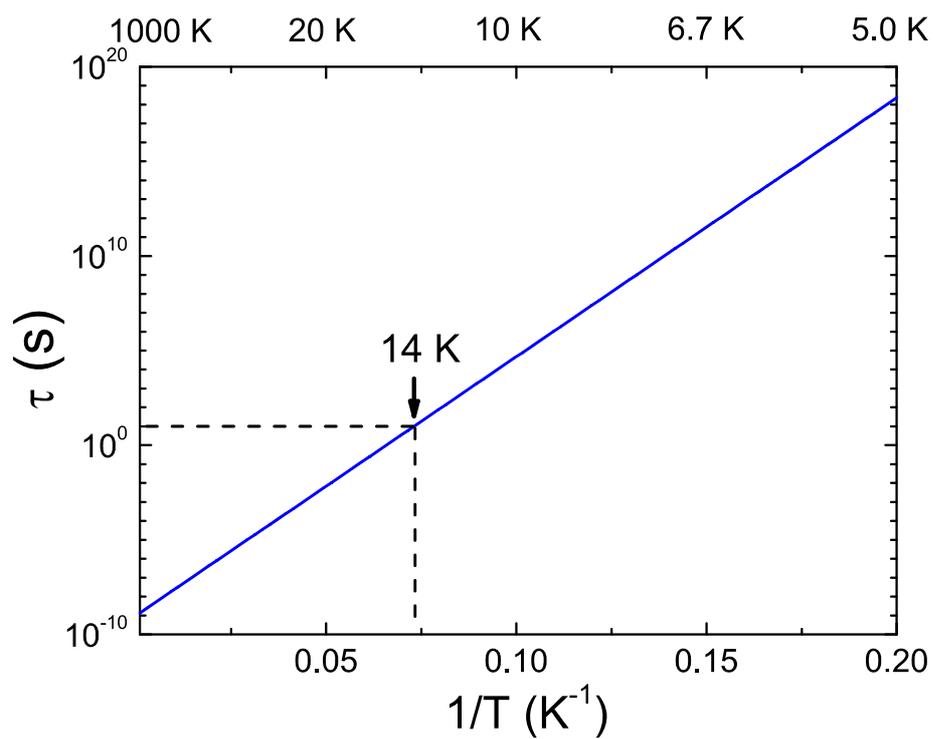

Fig. 3, Petracic et al.

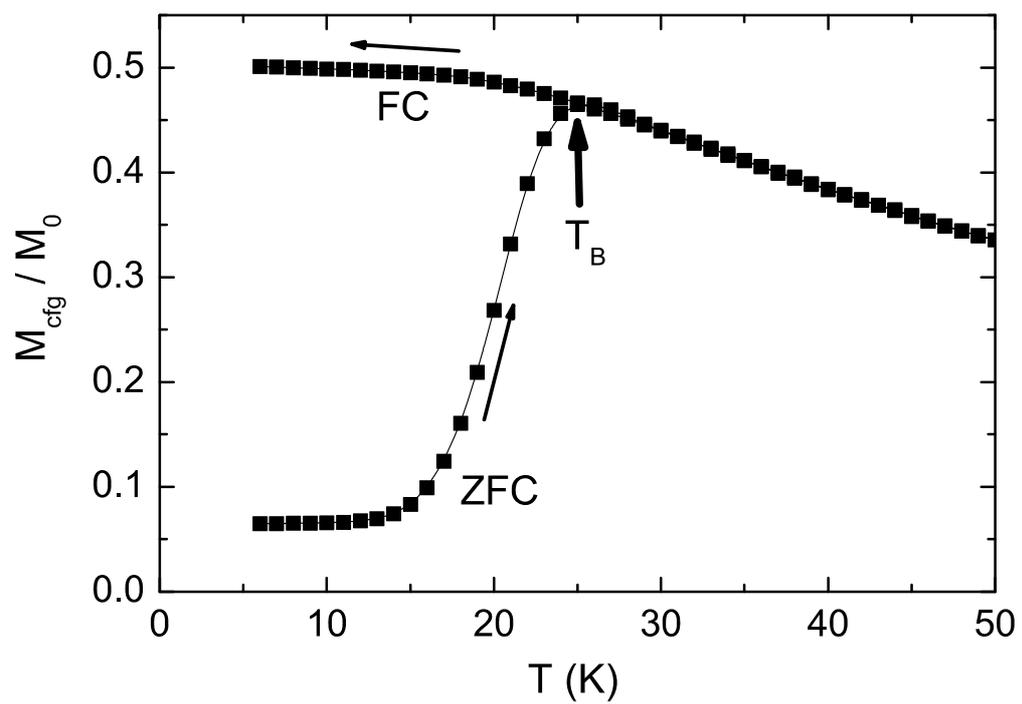

Fig. 4, Petracic et al.

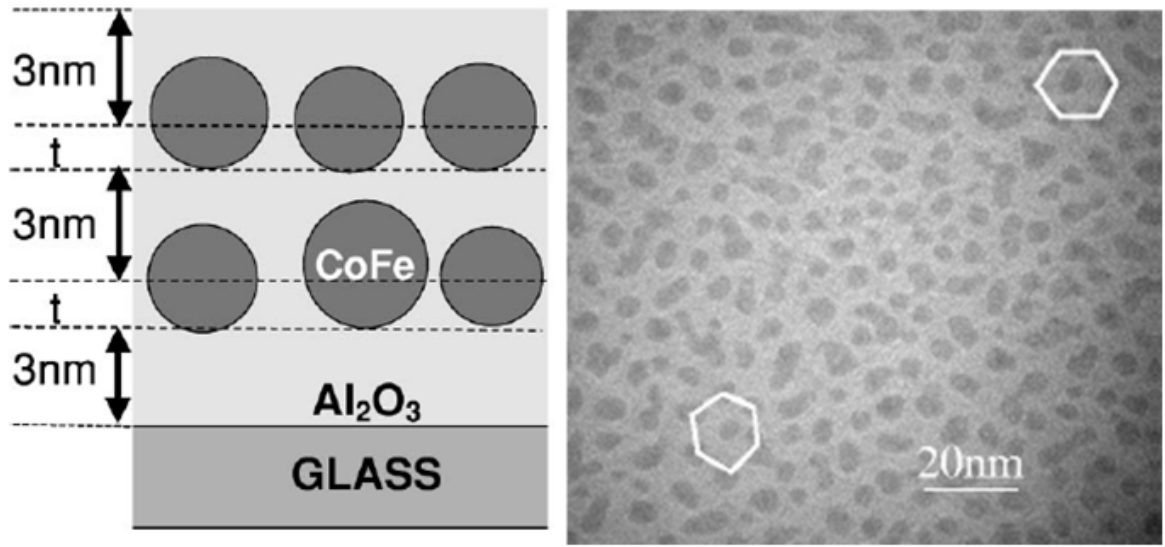

Fig. 5, Petracic et al.

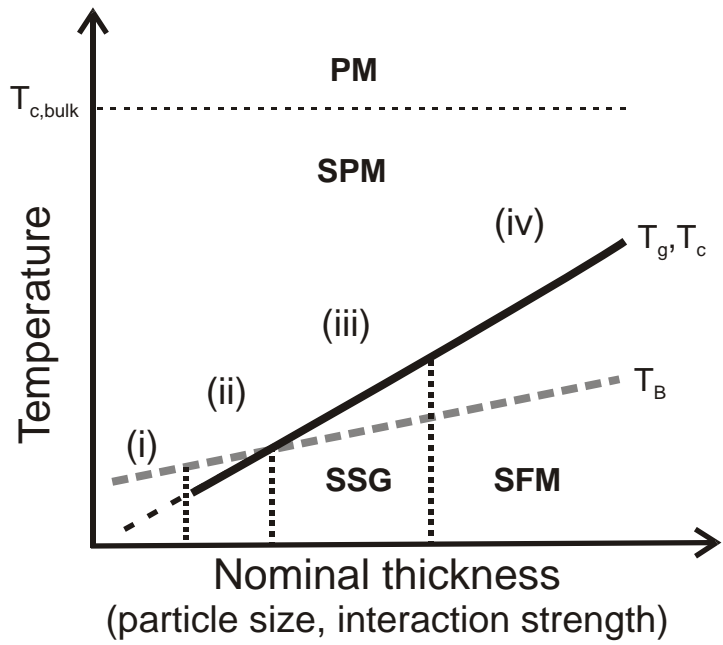

Fig. 6, Petracic et al.

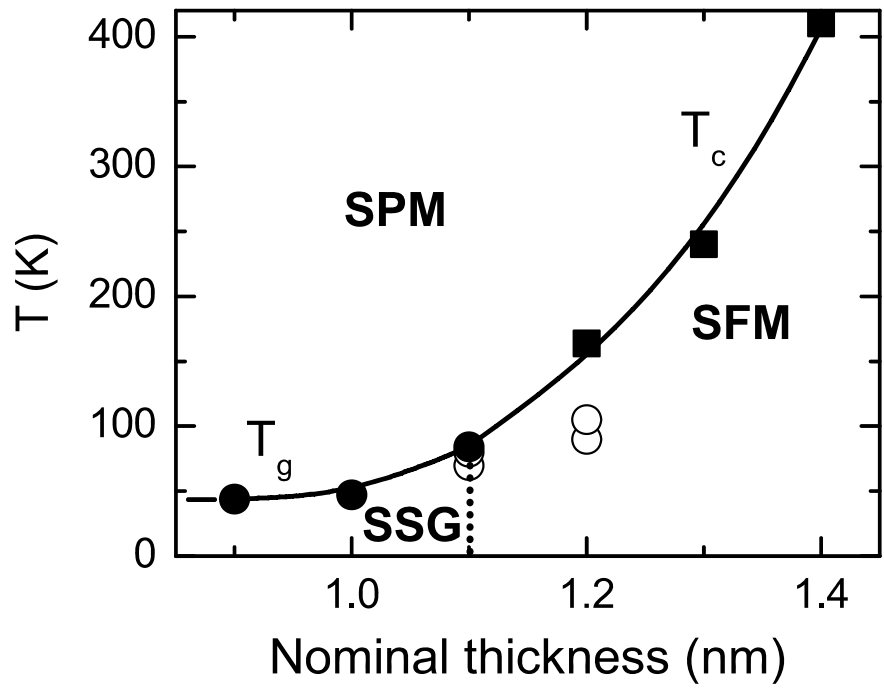

Fig. 7, Petracic et al.